\documentclass[10pt]{bmc_article}    

\usepackage{cite} 
\usepackage{url}  
\usepackage{ifthen}  
\usepackage{multicol}   
\usepackage[utf8]{inputenc} 
\usepackage{mathbbol,amssymb,latexsym,amsfonts,amsmath,amsthm}
\usepackage[pdftex]{graphicx}
\usepackage{color}
\usepackage{booktabs}
\usepackage{longtable}

\newboolean{publ}

\newenvironment{bmcformat}{\fussy\setboolean{publ}{true}}{\fussy}

\begin{document}
\begin{bmcformat}


  \title{Comparative analysis of module-based versus direct methods
    for reverse-engineering transcriptional regulatory networks}
 

\author{Tom Michoel\correspondingauthor$^{1,2}$%
  \email{Tom Michoel\correspondingauthor - tom.michoel@psb.vib-ugent.be}%
  \and
  Riet De Smet$^3$%
  \email{Riet De Smet - riet.desmet@esat.kuleuven.be}
  \and
  Anagha Joshi$^{1,2}$%
  \email{Anagha Joshi - anagha.joshi@psb.vib-ugent.be}
  \and
  Yves Van de Peer$^{1,2}$%
  \email{Yves Van de Peer - yves.vandepeer@psb.vib-ugent.be}
  and
  Kathleen Marchal$^{3}$%
  \email{Kathleen Marchal - kathleen.marchal@biw.kuleuven.be}
}


\address{%
  \iid(1)Department of Plant Systems Biology, VIB, Technologiepark
  927, B-9052 Gent, Belgium\newline
  \iid(2)Department of Molecular Genetics, Ghent University,
  Technologiepark  927, B-9052 Gent, Belgium\newline
  \iid(3)CMPG, Department Microbial and Molecular Systems,
  KU Leuven, Kasteelpark Arenberg 20, B-3001 Leuven, Belgium}%

\maketitle


\begin{abstract}
  \paragraph*{Background:} A myriad of methods to reverse-engineer
  transcriptional regulatory networks have been developed in recent
  years. Direct methods directly reconstruct a network of pairwise
  regulatory interactions while module-based methods predict a set of
  regulators for modules of coexpressed genes treated as a single
  unit. To date, there has been no systematic comparison of the
  relative strengths and weaknesses of both types of methods.
      
  \paragraph*{Results:} We have compared a recently developed
  module-based algorithm, LeMoNe (Learning Module Networks), to a
  mutual information based direct algorithm, CLR (Context Likelihood
  of Relatedness), using benchmark expression data and databases of
  known transcriptional regulatory interactions for
  \textit{Escherichia coli} and \textit{Saccharomyces cerevisiae}.  A
  global comparison using recall versus precision curves hides the
  topologically distinct nature of the inferred networks and is not
  informative about the specific subtasks for which each method is
  most suited.  Analysis of the degree distributions and a regulator
  specific comparison show that CLR is `regulator-centric', making
  true predictions for a higher number of regulators, while LeMoNe is
  `target-centric', recovering a higher number of known targets for
  fewer regulators, with limited overlap in the predicted interactions
  between both methods.  Detailed biological examples in \textit{E.
    coli} and \textit{S.  cerevisiae} are used to illustrate these
  differences and to prove that each method is able to infer parts of
  the network where the other fails.  Biological validation of the
  inferred networks cautions against over-interpreting recall and
  precision values computed using incomplete reference networks.

  \paragraph*{Conclusions:} Our results indicate that module-based and
  direct methods retrieve largely distinct parts of the underlying
  transcriptional regulatory networks. The choice of algorithm should
  therefore be based on the particular biological problem of interest
  and not on global metrics which cannot be transferred between
  organisms. The development of sound statistical methods for
  integrating the predictions of different reverse-engineering
  strategies emerges as an important challenge for future research.
\end{abstract}

\ifthenelse{\boolean{publ}}{\begin{multicols}{2}}{}


\section*{Background}

Due to the success of microarray technology, the available data on the
transcriptional regulatory networks of different organisms has grown
exponentially. In order to explore these data to the maximum, a myriad
of methods to reverse-engineer or reconstruct transcriptional
regulatory networks from microarray data have been developed in the
past few years. In general, the scientific community has mainly
focused on the overall performance of newly developed methods in
reconstructing the known network of certain model organisms as
compared to a reference network, measuring algorithmic performance
with standard measures such as recall and precision. Less attention
has been paid to what extent conceptually different approaches differ
in the networks they infer.  Nonetheless, in order to get a better
understanding of the systems studied it is also important to
understand which specific problems can be tackled using a certain
method, irrespective of the overall performance of the different
methods.

Broadly speaking we can distinguish between two classes of methods for
reverse-engineering transcriptional regulatory networks from gene
expression data which differ vastly in how they approach the network
inference problem. Direct methods infer individual regulator-target
interactions using a pairwise correlation measure between the
expression profiles of a transcription factor and its putative targets
\cite{basso2005,faith2007}. Module-based methods assume a modular
structure of the transcriptional regulatory network
\cite{segal2003,ihmels2002,bonneau2006}, with genes subject to the
same regulatory input being organized in coexpression modules.

While different direct methods have been compared to each other in the
past \cite{faith2007,soranzo2007,zampieri2008b}, no systematic
comparison between direct and module-based methods has been undertaken
so far. In this study we perform such a comparison using a
representative method from each class. The CLR (Context Likelihood of
Relatedness) algorithm \cite{faith2007} considers all possible
pairwise regulator-target interactions and scores these interactions
based on the mutual information of their expression profiles as
compared to an interaction specific background distribution. It has
been shown to outperform other direct methods \cite{faith2007}. The
LeMoNe (Learning Module Networks) algorithm \cite{joshi2008} uses
probabilistic, ensemble-based optimization techniques
\cite{joshi2007,joshi2008} to infer high-quality module networks
\cite{segal2003}, where genes are first partitioned into coexpression
modules and regulators are assigned to modules based on how well they
explain the condition-dependent expression behavior of the module. It
has been shown to outperform the original module network algorithm
\cite{joshi2008}.

We have compared both methods at increasing levels of detail using
public expression compendia for \textit{Escherichia coli}
\cite{faith2007} and \textit{Saccharomyces cerevisiae}
\cite{gasch2000}, two organisms for which relatively large databases
of known transcriptional regulatory interactions exist
\cite{salgado2006,balaji2006}.  We first use recall versus precision
curves to give a comparison of the global performance of both methods.
We then show that due to the different assumptions underlying both
methodologies, they infer topologically distinct networks with limited
overlap, even at equal performance thresholds.  To understand these
distinctions more completely, we examined in detail example subsystems
of the network which are well characterized, namely the chemotaxis and
flagellar system in \textit{E.  coli} and a respiratory module and a
membrane lipid and fatty acid metabolism module in \textit{S.
  cerevisiae}.  Biological validation of the inferred networks
cautions against over-interpreting recall and precision values
computed using incomplete reference networks.

\section*{Results and Discussion}

\subsection*{Global comparison using recall and precision}

\begin{figure*}[ht!]
  \centering
  \includegraphics[width=\linewidth]{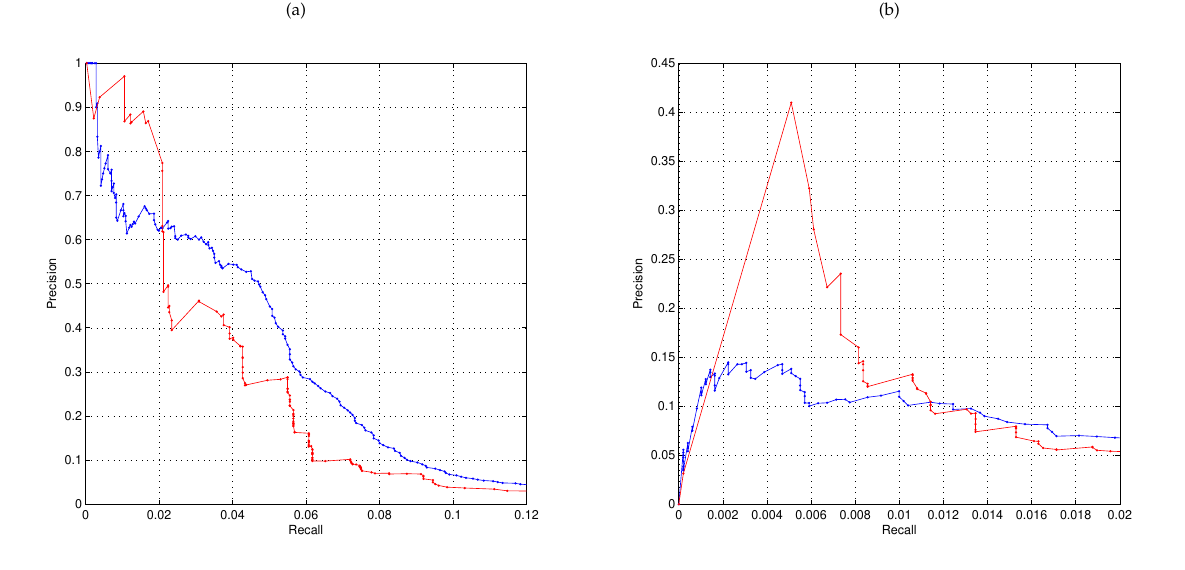}
  \caption{Recall versus precision curves for LeMoNe (red) and CLR
    (blue) for \textit{E. coli} (a) and \textit{S. cerevisiae} (b).
    Note the difference in scale between both organisms.}
  \label{fig:recall-precision}
\end{figure*}

The output of LeMoNe and CLR consists of a list of respectively ranked
regulator-module and ranked regulator-target interactions, scored
according to their statistical significance.  As a first, global,
comparison, we can therefore compute recall and precision with respect
to the given reference networks at different score cutoffs.  For CLR
we can directly compare the inferred network with the true network;
for LeMoNe we draw an edge between each regulator assigned to a module
and all genes in the module, thereby ignoring at this stage the extra
information present in the module structure. We computed recall and
precision as in \cite{faith2007}: if an edge is predicted between two
genes present but unconnected in the reference network it is counted
as a false positive. Figure \ref{fig:recall-precision} shows the
recall versus precision curves for both algorithms and both organisms.

Both algorithms succesfully prioritize true positive interactions,
especially in \textit{E. coli}: all curves go from a high precision,
low recall region to a low precision, high recall region. For CLR the
curves show a smooth course while for LeMoNe they are more
staircase-like. CLR scores individual interactions and as a result, in
the recall-precision curve interactions will be added one by one, but
interactions corresponding to a certain regulator will be dispersed
continuously throughout the recall-precision curve.  LeMoNe on the
other hand assigns a regulator to a module as a whole and all targets
belonging to the same module are added at the same time in the
recall-precision curve.  For a stringent threshold and subsequently a
low number of interactions inferred, the CLR network will cover few
interactions for many regulators while the LeMoNe network will
retrieve many interactions for few regulators.

At similar levels of precision, the recall in \textit{S.  cerevisiae}
is nearly an order of magnitude smaller than in \textit{E. coli}, in
line with previous studies \cite{zampieri2008}.  This is likely due to
the higher complexity of transcriptional regulation in \textit{S.
  cerevisiae} with a higher degree of combinatorial regulation and
posttranscriptional control, and consequently a lower degree of
correlation in expression between transcription factors and their
targets.

A simple `area under the curve' measurement would suggest that CLR
performs slightly better in the prokaryote \textit{E.  coli} and
LeMoNe in the eukaryote \textit{S.  cerevisiae}.  However, as we will
show below, both algorithms infer complementary information in both
organisms.

\subsection*{Topological distinctions between inferred networks}

\begin{figure*}[ht!]
  \centering
  \includegraphics[width=\linewidth]{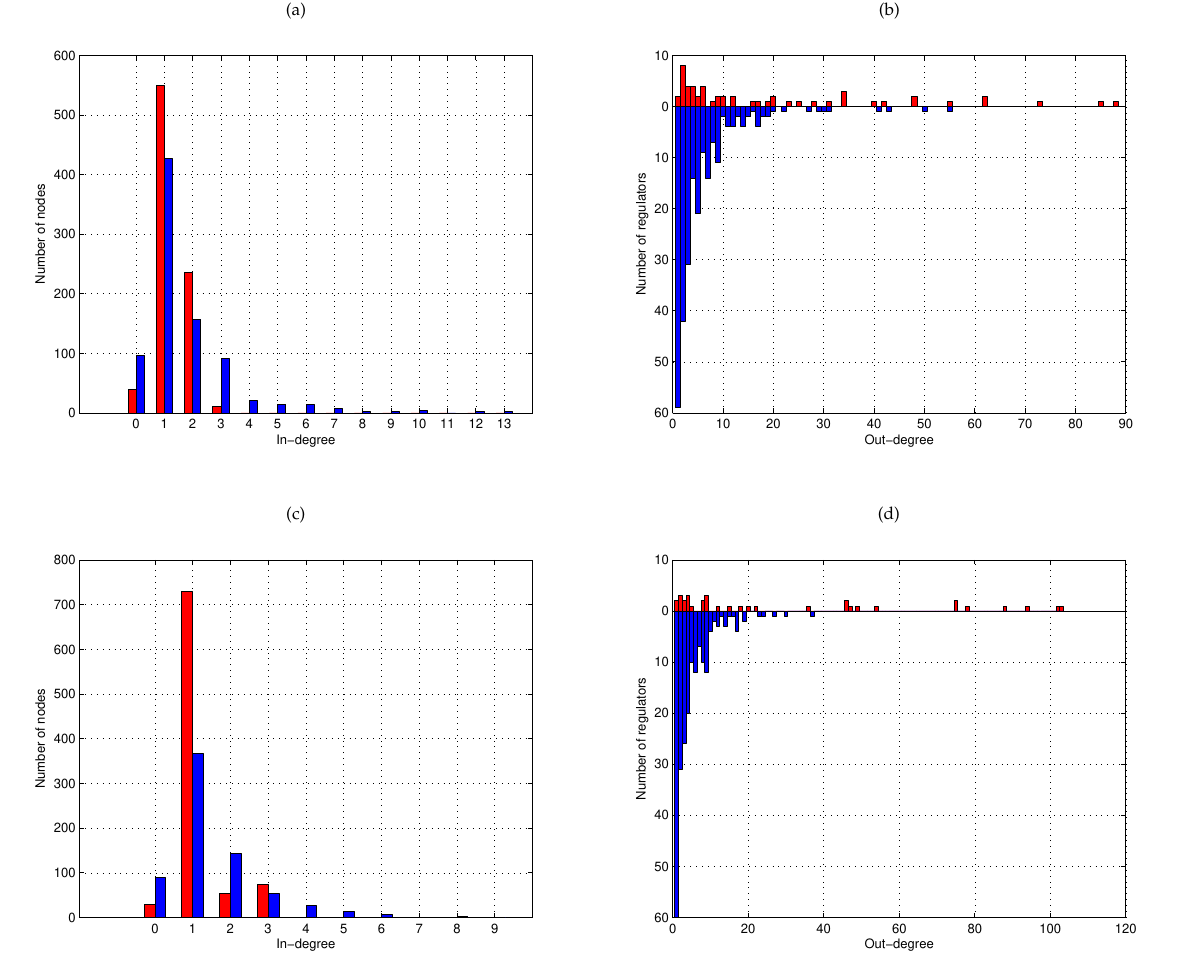}
  \caption{\textbf{(a)} \textit{E. coli} in-degree distribution for
    LeMoNe (red) and CLR (blue) at 30\% precision threshold.
    \textbf{(b)} \textit{E. coli} out-degree distribution for LeMoNe
    (red) and CLR (blue) at 30\% precision threshold. \textbf{(c)}
    \textit{S. cerevisiae} in-degree distribution for LeMoNe (red) and
    CLR (blue) at first 1070 predictions. \textbf{(d)} \textit{S.
      cerevisiae} out-degree distribution for LeMoNe (red) and CLR
    (blue) at first 1070 predictions.}
  \label{fig:degree-dist}
\end{figure*}

As explained in the previous section, due to how interactions are
scored, direct and module-based methods will infer different kinds of
networks at stringent precision thresholds. For \textit{E.  coli}, we
compared the LeMoNe and CLR networks at a 30\% precision threshold
where both networks have nearly equal recall and precision (see Figure
\ref{fig:recall-precision}). The LeMoNe network consists of 53
regulators assigned to 62 modules for a total of 1079 predicted
interactions; 594 of these interactions are between genes in
RegulonDB, with a precision of 29\%.  The corresponding CLR network
contains 1422 predicted interactions for 242 regulators; 597 of these
interactions are between genes in RegulonDB, with a precision of 30\%.
51 out of 53 LeMoNe regulators are also present in the CLR network,
but only 277 interactions are predicted in both networks. For
\textit{S. cerevisiae}, there is no `natural' point on the recall
versus precision curve to compare both networks.  We therefore
compared CLR and LeMoNe at the first 1070 predicted interactions. This
number is chosen to give comparably sized networks as in \textit{E.
  coli} and ensure that the ranked list of LeMoNe interactions is not
cut off in the middle of one module. The cutoff of the first 1070
interactions corresponds to precision values of respectively 16\% and
10\% for LeMoNe and CLR (cfr. Figure \ref{fig:recall-precision}).  The
LeMoNe network consists of 34 regulators assigned to 39 modules
containing 867 genes, while the CLR network contains 214 regulators;
28 regulators are present in both networks, yet only 75 interactions
are common.

The networks inferred by LeMoNe and CLR are topologically very
distinct (see Supplementary Figures S1 to S4). This distinction can be
quantified by their in- and out-degree distributions (Figure
\ref{fig:degree-dist}). The in-degree is the number of regulators
assigned to a certain target gene and the in-degree distribution
counts for each value $k$ the number of targets with in-degree $k$.
Likewise, the out-degree is the number of targets assigned to a
certain regulator and the out-degree distribution counts for each
value $k$ the number of regulators with out-degree $k$. CLR infers for
each regulator only the most significant targets.  As a result, the
out-degree distribution is skewed to the left, with the majority of
regulators having only few targets.  The in-degree distribution on the
other hand has a long tail of genes assigned to many different
regulators.  LeMoNe infers for each module the most significant
regulators, resulting in opposite characteristics of the degree
distributions. The in-degree distribution has no tail since for most
modules at most 2 significant regulators are identified. The
out-degree distribution on the other hand has a long tail since each
regulator assignment involves a whole module of genes. For these
reasons, we say that CLR is `regulator-centric' and LeMoNe is
`target-centric'.

\subsection*{Regulator specific comparison}

\begin{figure*}[ht!]
  \centering
  \includegraphics[width=\linewidth]{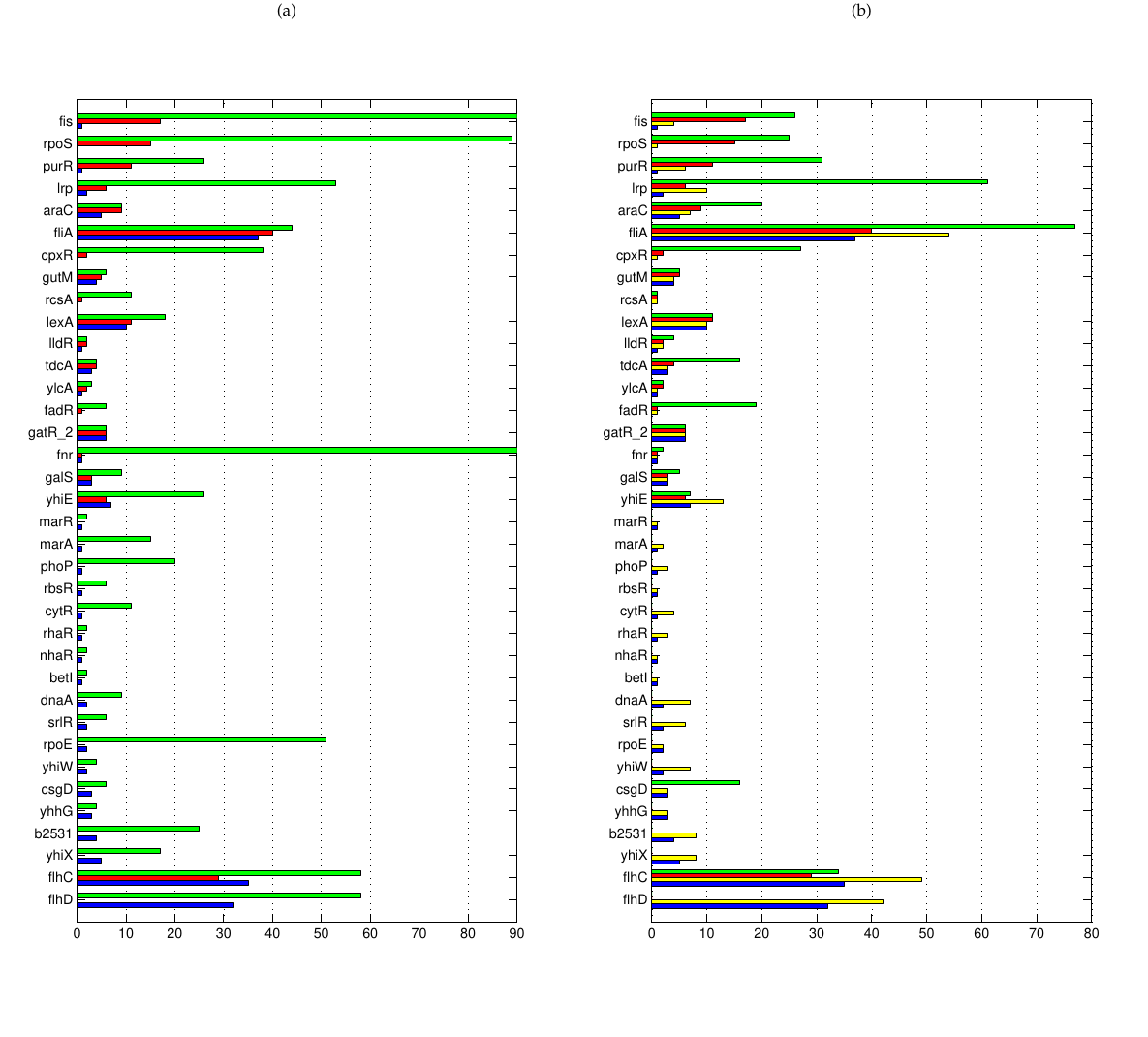}
  \caption{ For each regulator in \textit{E.  coli} with known
    interactions inferred: \textbf{(a)} the number of interactions in
    the reference network (green) and the number of true positives in
    LeMoNe (red) and CLR (blue); \textbf{(b)} the number of
    interactions inferred (green) and the number of true positives
    (red) in LeMoNe, and the number of interactions inferred (yellow)
    and the number of true positives (blue) in CLR. LeMoNe and CLR
    networks are both at 30\% precision threshold.  Regulators are
    sorted by the difference
    $\text{TP}_{\text{LeMoNe}}-\text{TP}_{\text{CLR}}$. The total
    number of true positives is 171 for LeMoNe and 180 for CLR.  For
    clarity, the $x$-axis in (a) is truncated, the true number of
    targets for Fis and Fnr is respectively 111 and 173.  The number
    of interactions inferred only counts targets that belong to the
    reference network.}
  \label{fig:regulators-ecoli}
\end{figure*}

\begin{figure*}[ht!]
  \centering
  \includegraphics[width=\linewidth]{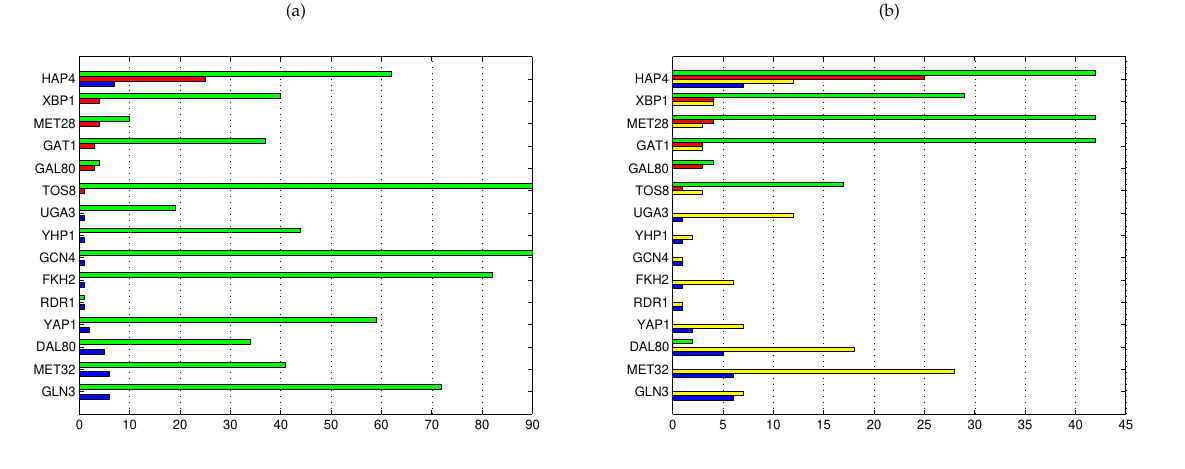}
  \caption{For each regulator in \textit{S. cerevisiae} with known
    interactions inferred: \textbf{(a)} the number of interactions in
    the reference network (green) and the number of true positives in
    LeMoNe (red) and CLR (blue); \textbf{(b)} the number of
    interactions inferred (green) and the number of true positives
    (red) in LeMoNe, and the number of interactions inferred (yellow)
    and the number of true positives (blue) in CLR.  LeMoNe and CLR
    networks are both cut off at the first 1070 predictions.
    Regulators are sorted by the difference
    $\text{TP}_{\text{LeMoNe}}-\text{TP}_{\text{CLR}}$.  The total
    number of true positives is 40 for LeMoNe and 31 for CLR. For
    clarity, the $x$-axis in (a) is truncated, the true number of
    targets for GCN4 is 120. The number of interactions inferred only
    counts targets that belong to the reference network.}
  \label{fig:regulators-yeast}
\end{figure*}

We make a further comparison of the two methods, focusing on how they
differ in the type of regulators they assign.  We compared again the
30\% precision networks for \textit{E.  coli} and the networks of
first 1070 interactions for \textit{S. cerevisiae}.

For both methods, a large fraction of the regulators for which known
targets are inferred are autoregulators. For \textit{E.  coli}, LeMoNe
and CLR have respectively 19 and 32 regulators with at least one true
positive; 15/19 (79\%) and 27/32 (84\%) are known autoregulators,
while the fraction of autoregulators in the total reference network is
95/150 (63\%). For \textit{S. cerevisiae}, LeMoNe and CLR have
respectively 6 and 10 regulators with at least one true positive; 5/6
(83\%) and 5/10 (50\%) are known autoregulators, while the fraction of
autoregulators in the total reference network is 79/171 (46\%).  The
abundance of autoregulators is not surprising since autoregulation is
a simple mechanism by which the expression profile of a regulator and
its targets can be correlated.

In LeMoNe, we get as additional information whether a predicted
regulator is positively or negatively correlated with its target
module and RegulonDB, the reference network for \textit{E. coli},
contains the activation or repression sign for many interactions.
However, although theoretically possible, we could not detect
biologically relevant patterns of anticorrelation, in line with
previous studies \cite{herrgaard2003}.  Even though the assumption of
anticorrelation seems intuitively plausible in case of repressors, it
is a too simplistic representation of reality. Indeed LeMoNe and CLR
both find many targets of mainly autorepressors (e.g. LexA, PurR, LldR
and GalS), but they all were positively instead of negatively
correlated with their targets.  This can be explained by the fact that
the activity of such autorepressors is dependent upon the presence of
corepressing signals. In the absence of the corepressing signal the
repressor is active, limiting its own production as well as that of
its target genes. In presence of the corepressing signal the
repressors are inactive, which enables the production of both inactive
repressor gene and its targets \cite{mangan2006,michel2005,meng1990}.

In \textit{E. coli}, regulators for which the module-based and direct
methods differ in performance are in line with the topological
distinctions.  CLR is better at inferring interactions for regulators
that are known to regulate just one or a few operons (e.g. BetI, CsgD,
DnaA, MarA, Yhhg, see Figure \ref{fig:regulators-ecoli}). These
operons are found with a relatively high rank in the CLR network since
their regulators often belong themselves to the operons and are thus
by definition tightly coexpressed with their targets. The clustering
method employed by LeMoNe appears to be too coarse grained to identify
these operons individually, since they are mostly part of larger
clusters. LeMoNe on the other hand is superior at inferring
interactions for regulators that are known to regulate larger
regulons, such as Fis, LexA, PurR, and RpoS, for which the level of
coexpression is not as high as the one observed within a single operon
(see Figure \ref{fig:regulators-ecoli}). In \textit{S.  cerevisiae},
there is no operonic structure and hence the `operon regulators'
acurately identified by CLR are absent. Figure
\ref{fig:regulators-yeast} show however that the regulators for which
LeMoNe and CLR infer known targets are still very distinct, but there
appears to be no general biological reason underlying these
differences.

\subsection*{Biological validation of inferred networks}

Due to the lack of a negative gold standard, we have denoted in the
previous analysis an edge as being false positive if both regulator
and target are present but not connected in the reference network (the
positive gold standard). Since the coverage of these reference
networks is still very incomplete, it is likely that the number of
false positives is overestimated.  Moreover, about half of the
regulators in \textit{E.  coli} and \textit{S. cerevisiae} are not
present in the reference network and their predicted interactions are
thus never evaluated.  

In \cite{faith2007}, it was already shown that new predictions made by
CLR in \textit{E. coli} could be validated experimentally. Here we
have performed an in-depth biological validation of the 30\% precision
module network inferred by LeMoNe. To biologically validate the
obtained regulator-module assignments, we calculated for all modules
functional enrichment scores \cite{keseler2005} and enrichment in
targets of previously annotated regulators \cite{salgado2006}.  Table
\ref{tab:lemone-ecoli} shows that in nearly all cases the module is
enriched in known targets of the predicted regulator (column 4) or at
least involved in the same biological function (column 6). In several
cases the predicted regulator is the one which has the best target
enrichment $p$-value.  Nearly half of the regulators are putative
regulators without any currently known targets, and these assignments
cannot be validated.  However, many of the correctly predicted
regulators involve neighbor regulators \cite{hershberg2005} (Table
\ref{tab:lemone-ecoli}, column 7), i.e.  regulators colocalized with
their targets on the genome. It has been suggested that many of the
putative regulators in \textit{E.  coli} constitute such neighbor
regulators \cite{price2008}. Hence this feature of gene neighborhood
can be used to attach additional significance to the high-scoring
predictions for uncharacterized regulators.  One of the advantages of
a module-based approach is the fact that if a certain module contains
several known targets of the assigned regulators, the rest of the
unknown targets in this module can be considered high confidence
predictions for that regulator. This is illustrated in Supplementary
Table S1, where we list several predictions for 10 different modules
which could be confirmed by a thorough literature search.

Module network predictions in \textit{S. cerevisiae} have been
experimentally validated in \cite{segal2003} and functionally analysed
in \cite{segal2003,joshi2008}. For further validation we compared the
CLR and LeMoNe networks to the YEASTRACT database \cite{teixeira2006}.
This database contains most of the interactions in the reference
network we use here \cite{balaji2006}. In addition it also contains
targets inferred by transcription factor deletion microarray
experiments.  The number of true positives for the LeMoNe network cut
off at the first 1070 predictions increases from 40 (precision 16\%)
in the reference network to 55 (precision 24\%) with respect to
YEASTRACT.  For the CLR network cut off at the first 1070 predictions,
the number of true positives increases from 31 (precision 10\%) in the
reference network to 48 (precision 12\%) with respect to YEASTRACT.

Biological validation of inferred networks is tedious and does not
provide an easy alternative to the automatic estimation of true and
false positives using an established reference network. The results of
this section do show however that many `false positives' with respect
to an incomplete network are actually true positives when additional
information is taken into account and that recall versus precision
plots such as in Figure \ref{fig:recall-precision} have to be
interpreted with caution.

\subsection*{The chemotaxis and flagellar system in
  \textit{Escherichia coli}}

\begin{figure*}[ht!]
  \centering
  \includegraphics[width=\linewidth]{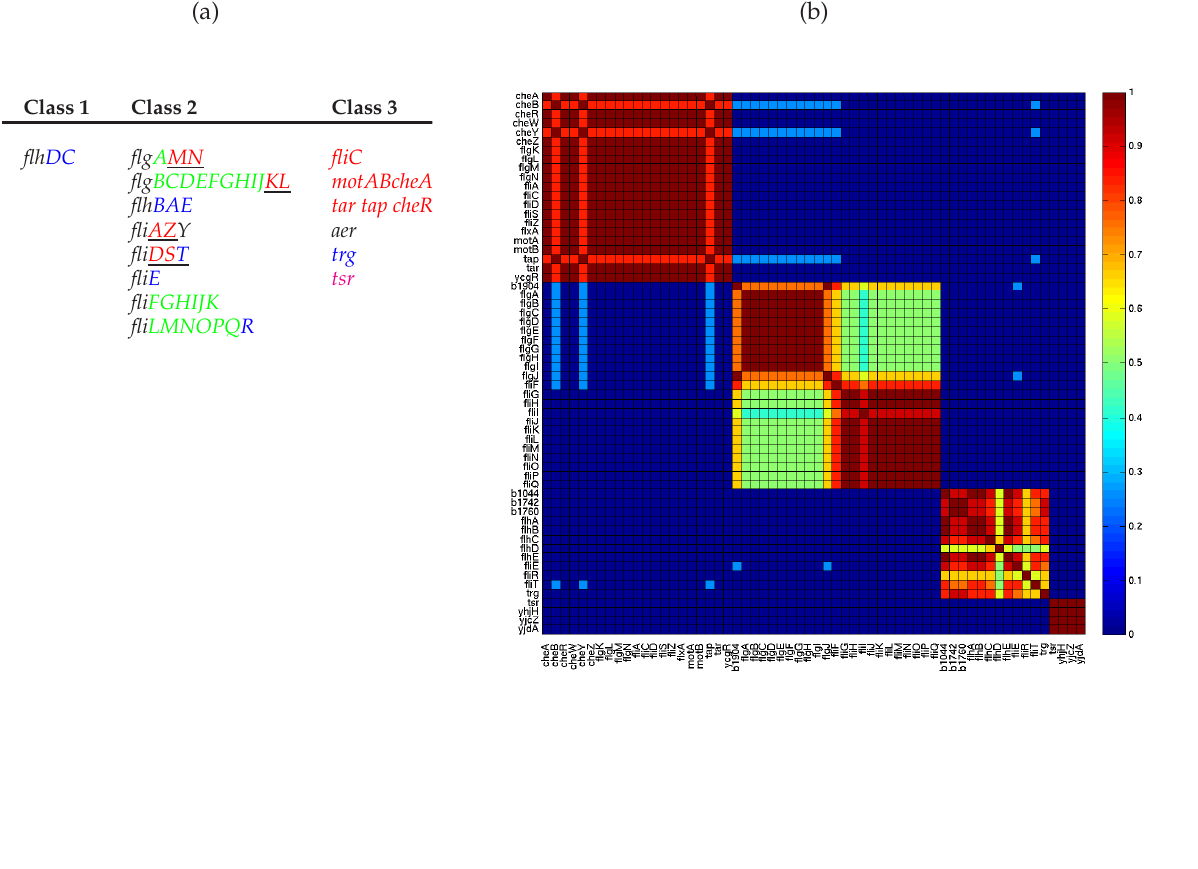}
  \vspace*{-25mm}
  \caption{\textbf{(a)} Operons encoding the proteins of the
    chemotaxis and flagellar system in \textit{E. coli}. The
    underlined genes belong to operons activated by FlhDC but have
    additional promoters activated by FliA. They are expressed
    partially as class 2 genes and fully as class 3 genes. Table and
    data after \cite{berg2003}.  Genes belonging to module 12 are
    indicated in red, to module 18 in green, to module 24 in blue and
    to module 45 in magenta.  \textbf{(b)} Pairwise clustering
    frequencies in the LeMoNe clustering ensemble
    \cite{joshi2007,joshi2008} for the flagella genes. Each row/column
    corresponds to a gene in one of the flagella modules and the heat
    map value at position $(i,j)$ is the frequency with which gene $i$
    and $j$ cluster together. The blocks along the diagonal correspond
    to respectively module 12, 18, 24 and 45.  In module 24, it can be
    seen that the coclustering frequencies of \textit{flhD} with the
    other members is rather low, indicating a weaker degree of
    coexpression. See also Supplementary Figure S5.}
  \label{fig:flagella}
\end{figure*}

Our analysis has shown that at equal levels of recall and precision,
LeMoNe predicts interactions for fewer regulators but with higher
coverage per regulator while CLR predicts fewer interactions per
regulator but for more regulators. It is instructive to analyse in
detail the implications of these differences for subsystems of the
transcriptional regulatory network which are particularly well
perturbed in the data set. For \textit{E. coli}, we have taken a
closer look at the chemotaxis and flagellar system which forms a
complex and tightly regulated system.  It consists of the class 1
master operon \textit{flhDC}, 8 class 2 operons activated by the
complex FlhDC, and at least 6 class 3 operons activated by the sigma
factor FliA (Figure \ref{fig:flagella} (a)). The \textit{fliA} operon
belongs to class 2, positively regulates its own production and can
activate other class 2 operons as well \cite{berg2003}.

Four modules (12, 18, 24 and 45) in the module network are enriched in
flagellar functions.  Together they contain 60 genes of which 55 are
known flagellar genes.  The separation of flagellar genes in different
modules is strongly supported by the LeMoNe clustering (Figure
\ref{fig:flagella} (b)), suggesting the presence of condition-specific
regulation in the flagellar gene network, and corresponds to the
difference in regulatory input between different classes of flagellar
genes (Figure \ref{fig:flagella}, see also Supplementary Figure S5).
In the 30\% precision LeMoNe network, FliA is assigned to all four
modules and FlhC is correctly assigned to the class 2 modules 18 and
24 only. FlhD is not assigned with a score high enough to make the
threshold.

At the 30\% precision cutoff, LeMoNe and CLR agree for the majority of
predicted interactions for FliA and FlhC. In addition, CLR infers
several correct targets for FlhD.  The coexpression of FlhD with its
predicted targets is significantly lower than for FliA or FlhC. This
is evidenced for instance from the LeMoNe clustering (Figure
\ref{fig:flagella} (b)) or CLR mutual information values (data not
shown).  However, due to the regulator-centric viewpoint and the
`local' background correction method of CLR, these relatively weakly
coexpressed targets still get a significant mutual information
$z$-score and are thus part of the predicted network. In the
target-centric LeMoNe network, the potential assignment of FlhD to the
flagella modules is compared to the much better scoring assignments of
FliA and/or FlhC and therefore not deemed significant enough.  Hence
the regulator-centric CLR approach has the advantage to identify
significant targets for all three flagellar regulators, but does not
distinguish well between regulation by FlhDC and FliA due to the large
overlap in predicted targets. The target-centric LeMoNe approach on
the other hand has the advantage to infer detailed condition-specific
regulatory information through the division in distinct modules of the
flagellar genes, but only infers targets for FliA and FlhC.

\subsection*{The respiratory module and membrane lipid and fatty acid
  metabolism module in \textit{Saccharomyces cerevisiae}}

\begin{figure*}[ht!]
  \centering
  \includegraphics[width=\linewidth]{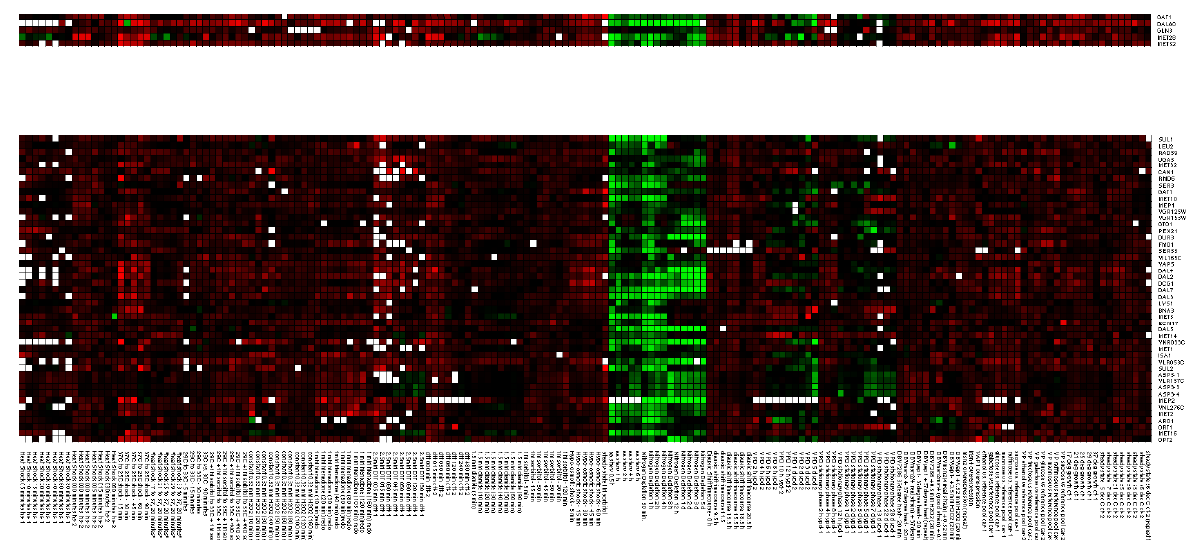}
  \caption{LeMoNe module 11 with genes (bottom) and predicted
    regulators (top) involved in the methionine pathway (regulated by
    Met28 and Met32) and the nitrogen catabolite repression system
    (regulated by Gat1, Dal80 and Gln3). The regulators are, from top
    to bottom: Gat1, predicted by LeMoNe, 1 target predicted by CLR;
    Dal80, 11 targets predicted by CLR; Gln3, 6 targets predicted by
    CLR; Met28, predicted by LeMoNe, 4 targets predicted by CLR;
    Met32, 23 targets predicted by CLR. The upregulated (green)
    conditions are all amino acid starvation or nitrogen depletion
    conditions.}
  \label{fig:yeast-mod11}
\end{figure*}

Despite the overall low performance on \textit{S. cerevisiae}, LeMoNe
and CLR both achieve good results on particular subsystems. The
advantage of a target-centric approach is well exhibited by the
respiratory system. This system is well perturbed in the data set and
clusters of respiratory genes are found repeatedly in it using various
approaches \cite{segal2003,chen2007,joshi2008}. LeMoNe module 7
contains 30 genes of which 23 are known respiratory genes. Hap4, a
global regulator of respiratory genes, is the most significant
regulator for this module and indeed 25 of its genes are known Hap4
targets. The pairwise correlation between Hap4 and its targets varies,
and since CLR scores all interactions individually, they are dispersed
throughout the ranked list of interactions. As a result, there are
only 12 predicted Hap4 targets (7 TP) in the first 1070 CLR
interactions (see also Figure \ref{fig:regulators-yeast}). Clearly,
the preliminary step of clustering genes into target modules was
necessary here to infer the complete Hap4 regulated module.

Another interesting example is given by LeMoNe module 11, a module of
47 genes involved in membrane lipid and fatty acid metabolism. The
four highest-ranked regulators by LeMoNe for this module (Gat1, Met28,
Met32 and Dal80) all have known targets in it. However, due to how
regulators are scored in LeMoNe, there are rarely more than two
significant regulators per module (see Figure \ref{fig:degree-dist}
(a) and (c)), and only the assignments of Gat1 (3 TP) and Met28 (4 TP)
are present in the network of the first 1070 LeMoNe interactions.  CLR
on the other hand finds the most significant targets for each
regulator individually and thus identifies correct targets from module
11 for the other regulators as well: Met28 (1 TP), Met32 (6 TP) and
Dal80 (6 TP).  For Gat1, CLR does not find true positives, however it
finds 5 TP in module 11 for a fifth regulator Gln3. Hence for this
module, the most complete information is retrieved by combining the
output of LeMoNe and CLR. The genes and predicted regulators of module
11 are mostly involved in 2 pathways, the methionine pathway
(regulated by Met28 and Met32) and the nitrogen catabolite repression
(NCR) system (regulated by Gat1, Dal80 and Gln3). Module 11 is
overexpressed in nitrogen depletion and amino acid starvation
conditions.  For NCR-sensitive genes it is known that they are not
activated when rich nitrogen sources are available, but get expressed
when only poor sources are left. A link between the methionine pathway
and nitrogen depletion, as predicted by LeMoNe through the clustering
and by CLR through the assignment of common targets to these
regulators, is not evident but appears to be confirmed by an ongoing
study \cite{mendesfereira2008}.

\subsection*{Conclusion}

In recent years, a wide variety of methods to reverse-engineer
transcriptional regulatory networks from microarray data have been
developed. Whereas the development of a new method mostly coincides
with a comparison in overall performance to all existing methods, so
far no in-depth study on how conceptual differences relate to
differences in the inferred networks have been made.  Here we
distinguished between two main approaches for reverse-engineering
transcriptional regulatory networks: the module-based approach and the
direct approach.  We compared a representative algorithm of each
approach (module based LeMoNe versus direct CLR) at several levels of
detail for two different organisms, the prokaryote \textit{E. coli}
and the eukaryote \textit{S. cerevisiae}. We have found that CLR is
`regulator-centric', making few but highly significant predictions for
a large number of regulators.  LeMoNe on the other hand is
`target-centric', identifying few but highly significant regulators
for a large number of genes grouped in coexpression modules.  Through
a regulator specific comparison and analysis of specific biological
subsystems, we have shown that at stringent significance cutoffs, the
conceptual differences in statistically scoring potential regulatory
interactions lead to topologically distinct inferred networks
containing different kinds of regulators and biological information.
Our results show that the choice of algorithm should be made primarily
based on whether the biological question under study falls within the
target-centric or regulator-centric viewpoint, and not on global
metrics which cannot be transferred between organisms.  Ideally,
several network inference strategies should be combined for the best
overall performance. It is an important challenge for future research
to develop sound statistical methods for optimally combining the
output of multiple, existing reverse-engineering algorithms.

\section*{Methods}

The \textit{E. coli} microarray data compendium \cite{faith2007}
contains expression profiles for 4345 genes under 189 different stress
conditions and genetic perturbations. We selected a subset of 1882
differentially expressed genes (standard deviation larger than 0.5)
and used a list of 316 known or putative transcription factors
\cite{salgado2006,keseler2005} to reconstruct regulatory networks.
LeMoNe \cite{joshi2008} (software available at
\url{http://bioinformatics.psb.ugent.be/software/details/LeMoNe})
identified 108 ensemble-averaged modules from 12 independent Gibbs
sampler runs, containing 1761 genes in total. It inferred a ranked
list of regulator-module edges from an ensemble of 10 regulatory
programs per module with 100 regulator samples per regulatory program
node (see \cite{joshi2008} for more details on the meaning of these
parameters). We applied CLR \cite{faith2007} (software available at
\url{http://gardnerlab.bu.edu/clr.html}) on the data for the 2084
selected genes (the union of the 1882 differentially expressed genes
and 316 candidate regulators) and kept all mutual information
$z$-scores between the 316 transcription factors and 1882 target
genes. As a reference network we used RegulonDB version 5.7
\cite{salgado2006}, a database of 4840 known transcriptional
interactions in \textit{E.  coli} between 167 transcription factors
and 1693 genes. Recall values are computed with respect to RegulonDB
restricted to the subset of 2084 genes. This subnetwork contains 3110
edges between 150 transcription factors and 1053 genes. We used EcoCyc
\cite{keseler2005} to compute functional enrichment of modules. Target
and functional enrichment in Table \ref{tab:lemone-ecoli} were
computed using a cumulative hypergeometric distribution, Bonferroni
corrected for multiple testing, with confidence level 95\%.

The \textit{S. cerevisiae} microarray data compendium \cite{gasch2000}
contains expression profiles for 6153 genes in 173 different stress
conditions. We used the same subset of 2355 differentially expressed
genes, including a list of 321 potential regulators, as used in
previous studies of this data set \cite{segal2003,joshi2008}. LeMoNe
was run with the same settings as for \textit{E. coli} and inferred 55
ensemble-averaged modules containing 1075 genes.  As reference network
we used a network recently compiled from the results of genetic,
biochemical and ChIP-chip experiments \cite{balaji2006}. It contains
11785 interactions between 154 transcription factors and 4047 genes.
After restriction to the subset of 2355 differentially expressed
genes, it contains 4513 interactions between 133 transcription factors
and 1628 genes.  The YEASTRACT \cite{teixeira2006} database contains
30979 transcriptional interactions in \textit{S. cerevisiae} between
171 transcription factors and 5727 genes. After restriction to the
subset of 2355 differentially expressed genes, it contains 12021
interactions between 137 transcription factors and 2182 genes.

\section*{Authors contributions}

TM developed software, analyzed data and wrote the manuscript. RDS
analyzed data and wrote the manuscript. AJ developed software and
analyzed data. YVdP supervised the study. KM wrote the manuscript and
supervised the study. All authors read and approved the final
manuscript.

\section*{Acknowledgements}
  \ifthenelse{\boolean{publ}}{\small}{}

  RDS is a research assistant of the IWT. AJ is supported by an
  Early-Stage Marie Curie Fellowship.  This work is supported by 1)
  Research Council KUL: GOA AMBioRICS,GOA/08/011, CoE EF/05/007
  SymBioSys, 2) FWO: projects G.0318.05, 3) IWT: SBO-BioFrame, 4) IUAP
  P6/25 (BioMaGNet).


{\ifthenelse{\boolean{publ}}{\footnotesize}{\small}
  \bibliographystyle{bmc_article}  

\begin{thebibliography}{10}
    \providecommand{\url}[1]{[#1]}
    \providecommand{\urlprefix}{}
    
  \bibitem{basso2005}
    Basso K, Margolin AA, Stolovitzky G, Klein U, Dalla-Favera R, Califano A:
    \textbf{Reverse engineering of regulatory networks in human B cells}.
    \emph{Nat Genet} 2005, \textbf{37}:382--390.
    
  \bibitem{faith2007}
    Faith JJ, Hayete B, Thaden JT, Mogno I, Wierzbowski J, Cottarel G, Kasif S,
    Collins JJ, Gardner TS: \textbf{Large-scale mapping and validation of
      \textit{Escherichia coli} transcriptional regulation from a compendium of
      expression profiles}. \emph{PLoS Biol} 2007, \textbf{5}:e8.
    
  \bibitem{segal2003}
    Segal E, Shapira M, Regev A, Pe'er D, Botstein D, Koller D, Friedman N:
    \textbf{Module networks: identifying regulatory modules and their
      condition-specific regulators from gene expression data}. \emph{Nat Genet}
    2003, \textbf{34}:166--167.
    
  \bibitem{ihmels2002}
    Ihmels J, Friedlander G, Bergmann S, Sarig O, Ziv Y, Barkai N:
    \textbf{Revealing modular organization in the yeast transcriptional network}.
    \emph{Nat Genet} 2002, \textbf{31}:370--377.

  \bibitem{bonneau2006}
    Bonneau R, Reiss DJ, Shannon P, Facciotti M, Hood L, Baliga NS, Thorsson V:
    \textbf{The {Inferelator}: an algorithm for learning parsimonious regulatory
      networks from systems-biology data sets \textit{de novo}}. \emph{Genome Biol}
    2006, \textbf{7}:R36.

  \bibitem{soranzo2007}
    Soranzo N, Bianconi G, Altafini C: \textbf{Comparing association network
      algorithms for reverse engineering of large scale gene regulatory networks:
      synthetic versus real data}. \emph{Bioinformatics} 2007,
    \textbf{23}:1640--1647.
    
  \bibitem{zampieri2008b}
    Zampieri M, Soranzo N, Bianchini D, Altafini C: \textbf{Origin of co-expression
      patterns in \textit{E. coli} and \textit{S. cerevisiae} emerging from reverse
      engineering algorithms}. \emph{PLoS One} 2008, \textbf{3}:e2981.
    
  \bibitem{joshi2008}
    Joshi A, De~Smet R, Marchal K, Van~de Peer Y, Michoel T: \textbf{Module
      networks revisited: computational assessment and prioritization of model
      predictions}. \emph{Bioinformatics} 2009, \textbf{25}:490--496.

  \bibitem{joshi2007}
    Joshi A, Van~de Peer Y, Michoel T: \textbf{Analysis of a {Gibbs} sampler for
      model based clustering of gene expression data}. \emph{Bioinformatics} 2008,
    \textbf{24}(2):176--183.
    
  \bibitem{gasch2000}
    Gasch AP, Spellman PT, Kao CM, Carmel-Harel O, Eisen MB, Storz G, Botstein D,
    Brown PO: \textbf{Genomic expression programs in the response of yeast cells
      to environmental changes}. \emph{Mol Biol Cell} 2000, \textbf{11}:4241--4257.
    
  \bibitem{salgado2006}
    Salgado H, Gama-Castro S, Peralta-Gil M, Diaz-Peredo E, Sanchez-Solano F,
    Santos-Zavaleta A, Martinez-Flores I, Jimenez-Jacinto V, Bonavides-Martinez
    C, Segura-Salazar J, Martinez-Antonio A, Collado-Vides J: \textbf{{RegulonDB}
      (version 5.0): Escherichia coli {K-12} transcriptional regulatory network,
      operon organization, and growth conditions}. \emph{Nucleic Acids Res} 2006,
    \textbf{34}:D394--397.

  \bibitem{balaji2006}
    Balaji S, Madan~Babu M, Iyer LM, Luscombe NM, Aravind L: \textbf{Comprehensive
      analysis of combinatorial regulation using the transcriptional regulatory
      network of yeast}. \emph{J Mol Biol} 2006, \textbf{360}:213--227.

  \bibitem{zampieri2008}
    Zampieri M, Soranzo N, Altafini C: \textbf{Discerning static and causal
      interactions in genome-wide reverse engineering problems}.
    \emph{Bioinformatics} 2008, \textbf{24}:1510--1515.
    
  \bibitem{herrgaard2003}
    Herrg\aa{}rd MJ, Covert MW, Palsson Bo: \textbf{Reconciling gene expression
      data with known genome-scale regulatory network structures}. \emph{Genome
      Res} 2003, \textbf{13}:2423--2434.

  \bibitem{mangan2006}
    Mangan S, Itzkovitz S, Zaslaver A, Alon U: \textbf{The incoherent feed-forward
      loop accelerates the response-time of the \textit{gal} system of
      \textit{Escherichia coli}}. \emph{J Mol Biol} 2006, \textbf{356}:1073--1081.
    
  \bibitem{michel2005}
    Michel B: \textbf{After 30 years of study, the bacterial SOS response still
      surprises us}. \emph{PLoS Biol} 2005, \textbf{3}:e255.
    
  \bibitem{meng1990}
    Meng LM, Nygaard P: \textbf{Identification of hypoxanthine and guanine as the
      co-repressors for the purine regulon genes of Escherichia coli}. \emph{Mol
      Microbiol} 1990, \textbf{4}:2187--2192.
    
  \bibitem{keseler2005}
    Keseler IM, Collado-Vides J, Gama-Castro S, Ingraham J, Paley S, Paulsen IT,
    Peralta-Gil M, Karp PD: \textbf{{{E}co{C}yc: a comprehensive database
        resource for {E}scherichia coli}}. \emph{Nucleic Acids Res} 2005,
    \textbf{33}(Database issue):334--337.

  \bibitem{hershberg2005}
    Hershberg R, Yeger-Lotem E, Margalit H: \textbf{Chromosome organization is
      shaped by the transcription regulatory network}. \emph{Trends Genet} 2005,
    \textbf{21}:138--142.

  \bibitem{price2008}
    Price MN, Dehal PS, Arkin AP: \textbf{Horizontal gene transfer and the
      evolution of transcriptional regulation in \textit{Escherichia coli}}.
    \emph{Genome Biol} 2008, \textbf{9}:R4.

  \bibitem{teixeira2006}
    Teixeira M, Monteiro P, Jain P, Tenreiro S, Fernandes A, Mira N, Alenquer M,
    Freitas A, Oliveira A, S\`a-Correia I: \textbf{{{T}he
        {Y}{E}{A}{S}{T}{R}{A}{C}{T} database: a tool for the analysis of
        transcription regulatory associations in {S}accharomyces cerevisiae}}.
    \emph{Nucleic Acids Res.} 2006, \textbf{34}:D446--451.

  \bibitem{berg2003}
    Berg HC: \textbf{The rotary motor of bacterial flegella}. \emph{Annu Rev
      Biochem} 2003, \textbf{72}:19--54.
    
  \bibitem{chen2007}
    Chen G, Jensen S, Stoeckert C: \textbf{{{C}lustering of genes into regulons
        using integrated modeling-{C}{O}{G}{R}{I}{M}}}. \emph{Genome Biol.} 2007,
    \textbf{8}:R4.

  \bibitem{mendesfereira2008}
    Mendes-Fereira A, Barbosa C, del Olmo M, Mendes-Faia A, Le\~{a}o C:
    \textbf{Expression profile of genes involved in hydrogen sulphide liberation
      by {S}accharomyces cerevisiae grown under different nitrogen concentrations}
    2008. [Available from Nature Precedings doi:10.1038/npre.2008.2736.1].

\end{thebibliography}

}


\ifthenelse{\boolean{publ}}{\end{multicols}}{}

\begin{table}[ht!]
  \fontsize{6pt}{6.5pt}
  \fontfamily{\sfdefault}
  \selectfont
  \centering
  \begin{tabular}{lccccccl}
    Regulator & Module ID & Score & Target enrich. & Autoreg. & Pathway & Local & Function\\
    \hline\\
    gatR\_2 & 73 & 1912.98&  $**$&  &  $*$&  $**$&  carbon utilization $>$ carbon compounds\\
    gadE & 48 & 1844.50&  $**$&  $*$&  $*$&  $**$& adaptations $>$ pH\\
    gutM & 38 & 1807.24&  $**$&  $*$&  $*$&  $*$&  carbon utilization $>$ carbon compounds\\
    \textbf{ymfN} & \textbf{58} & 1749.11&  &  &  &  $*$&  \\
    \textbf{ymfN} & \textbf{33} & 1711.17&  &  &  &  $*$&  \\
    fliA & 12 & 1510.48&  $**$&  $*$&  $*$&  $**$&  motility, chemotaxis, energytaxis; flagella; 
    biosynthesis of flagellum\\
    rcsB & \textbf{62} & 1261.72&  &  &  $*$&  $*$&  biosynthesis of colanic acid (M antigen)\\
    fecI & 57 & 1200.77&  &  $*$&  $*$&  &  adaptations $>$ Fe aquisition\\
    gatR\_2 & 42 & 1176.55&  $**$&  &  $*$&  $**$&  carbon utilization $>$ carbon compounds\\
    \textbf{yahA} & 82 & 1171.92&  &  &  &  &  \\
    rcsA & 87 & 1151.97&  $**$&  $*$&  $*$&  &  biosynthesis of colanic acid (M antigen)\\
    lexA & 20 & 996.62&  $**$&  $*$&  $*$&  $*$&  SOS response; DNA repair; protection $>$ 
    radiation\\
    lldR & 65 & 976.84&  $**$&  $*$&  $*$&  $*$& energy metabolism; aerobic respiration\\
    fliA & 45 & 956.70&  $**$&  $*$&  $*$&  &   motility, chemotaxis, energytaxis\\
    fliA & 18 & 903.46&  $*$&  $*$&  $*$&  & biosynthesis of flagellum;
    motility, chemotaxis, energytaxis; flagella\\
    nac & 85 & 827.17&  &  $*$&  $*$&  &   nitrogen metabolism\\
    \textbf{yiaG} & 15 & 816.55&  &  &  &  &  \\
    \textbf{ydaK} & 23 & 815.75&  &  &  &  $**$&  \\
    \textbf{ydaK} & 154 & 805.22&  &  &  &  &  \\
    fnr & 23 & 798.27&  $*$&  $*$&  $*$&  $**$&  energy metabolism; anaerobic respiration; 
    membrane\\
    lrp & 5 & 777.80&  &  $*$&  $*$&  & biosynthesis of building blocks $>$ amino acids\\
    araC & 46 & 760.44&  $**$&  $*$&  $*$&  $**$&  carbon utilization $>$ carbon compounds\\
    appY & 50 & 748.75&  &  &  &  &  \\
    \textbf{yfiE} & \textbf{67} & 736.50&  &  &  &  &  \\
    \textbf{osmE} & 15 & 734.87&  &  &  &  &  \\
    lexA & 78 & 726.67&  $**$&  $*$&  $*$&  &  SOS response\\
    purR & 144 & 708.63&  &  $*$&  $*$&  &  \\
    uidR & 81 & 708.36&  &  $*$&  &  &  \\
    araC & 21 & 678.10&  $*$&  $*$&  $*$&  &   carbon utilization $>$ carbon compounds\\
    \textbf{yfeG} & 29 & 663.94&  &  &  &  &  \\
    \textbf{b1450} & \textbf{53} & 662.16&  &  &  &  &  \\
    flhC & 18 & 650.64&  $**$&  &  $*$&  &  biosynthesis of flagellum;
    motility, chemotaxis, energytaxis; flagella\\
    \textbf{ogrK} & 83 & 645.35&  &  &  &  &  \\
    fliA & 17 & 637.28&  &  $*$&  &  &  \\
    rpoS & 14 & 637.13&  $**$&  &  $*$&  $*$& adaptations $>$ osmotic pressure \\
    pdhR & \textbf{55} & 633.52&  &  $*$&  $*$&  &  energy metabolism; anaerobic respiration\\
    tdcA & 31 & 619.06&  $*$&  $*$&  $*$&  $*$&   threonine catabolism;  carbon utilization $>$ 
    amino acids\\
    \textbf{yebK} & \textbf{106} & 617.44&  &  &  &  &  \\
    araC & 56 & 608.17&  $**$&  $*$&  $*$&  &   carbon utilization $>$ carbon compounds\\
    csgD & 26 & 599.30&  &  $*$&  &  &  \\
    hycA & 66 & 596.27&  &  &  &  &  \\
    tdcR & 11 & 593.75&  &  &  $*$&  &  carbon utilization $>$ amino acids\\
    fliA & 24 & 593.05&  $*$&  $*$&  $*$&  &  flagella;  motility, chemotaxis, energytaxis; 
    biosynthesis of flagellum\\
    chbR & 24 & 590.31&  &  $*$&  &  &  \\
    hycA & 29 & 563.45&  &  &  &  $*$&  \\
    galS & 76 & 561.25&  $**$&  $*$&  $*$&  $**$&   carbon utilization $>$ carbon compounds\\
    \textbf{nlp} & 77 & 559.41&  &  &  &  &  \\
    \textbf{yfeC} & 119 & 549.33&  &  &  &  &  \\
    \textbf{b1506} & 36 & 548.33&  &  &  &  &  \\
    lrp & 10 & 528.90&  $*$&  $*$&  $*$&  &   biosynthesis of building blocks $>$ amino acids\\
    \textbf{cspB} & 37 & 527.86&  &  &  &  &  \\
    cusR & 68 & 515.56&  $**$&  $*$&  $*$&  $**$&  extrachromosomal $>$ transposon related\\
    \textbf{b1284} & 51 & 514.78&  &  &  &  &  \\
    nanR & 9 & 508.87&  &  &  &  &  \\
    \textbf{yohL} & 90 & 496.21&  &  &  &  &  \\
    lrp & 126 & 493.60&  &  $*$&  $*$&  &   biosynthesis of building blocks $>$ amino acids\\
    \textbf{yjjQ} & \textbf{179} & 491.02&  &  &  &  &  \\
    \textbf{yehV} & \textbf{63} & 483.29&  &  &  &  &  \\
    \textbf{ogrK} & 27 & 481.75&  &  &  &  &  \\
    slyA & 3 & 474.43&  &  &  &  &  \\
    \textbf{ydcN} & 16 & 467.66&  &  &  &  &  \\
    cpxR & 9 & 465.39&  $*$&  $*$&  $*$&  &  adaptations $>$ other (mechanical, nutritional, 
    oxidative stress)\\
    \textbf{yehV} & 34 & 451.77&  &  &  &  &  \\
    fruR & \textbf{63} & 449.25&  &  &  &  &  \\
    araC & 64 & 441.57&  $*$&  $*$&  $*$&  &   carbon utilization $>$ carbon compounds\\
    fis & 19 & 436.12&  $**$&  $*$&  $*$&  $*$&  information transfer $>$ RNA related $>$ tRNA\\
    fadR & 16 & 435.98&  $*$&  &  &  &  \\
    purR & 10 & 431.78&  $**$&  $*$&  $*$&  &   biosynthesis of building blocks $>$ nucleotides\\
    cadC & 37 & 429.32&  &  $*$&  &  &  \\
    fecI & 54 & 429.28&  &  $*$&  &  &  \\
    \textbf{rstA} & \textbf{102} & 428.94&  &  &  &  &  \\
    tdcR & \textbf{61} & 428.84&  &  &  &  &  \\
    flhC & 24 & 426.88&  $**$&  &  $*$&  $*$&  flagella; motility, chemotaxis, energytaxis; 
    biosynthesis of flagellum\\
  \end{tabular}
  \fontfamily{\familydefault}
  \selectfont
  \caption{Biological validation of the LeMoNe 30\% precision network for \textit{E. coli}. 
    Target enrichment: ($*$) module is enriched in known targets of the predicted 
    regulator,  ($**$) module is most enriched for predicted regulator. 
    Autoregulator: ($*$) regulator is an autoregulator. Pathway: ($*$) module is 
    enriched in the same function(s) as the regulator. Local: ($*$) regulator is in 
    the same operon as the module genes, ($**$) Transcription unit of regulator is adjacent 
    to transcription units of  the module genes. Function: enriched functions of the module.  
    Regulators in bold face are putative regulators without known targets; module IDs 
    in bold face consist only of uncharacterized genes.}
  \label{tab:lemone-ecoli}
\end{table}

\newpage


\section*{Supplementary information (1 table + 5 figures)}

\setcounter{figure}{0}
\renewcommand{\thefigure}{S\arabic{figure}}

\setcounter{table}{0}
\renewcommand{\thetable}{S\arabic{table}}

\begin{table}[h!]
  \centering
  \includegraphics[width=0.82\linewidth,viewport=100 160 480 700]{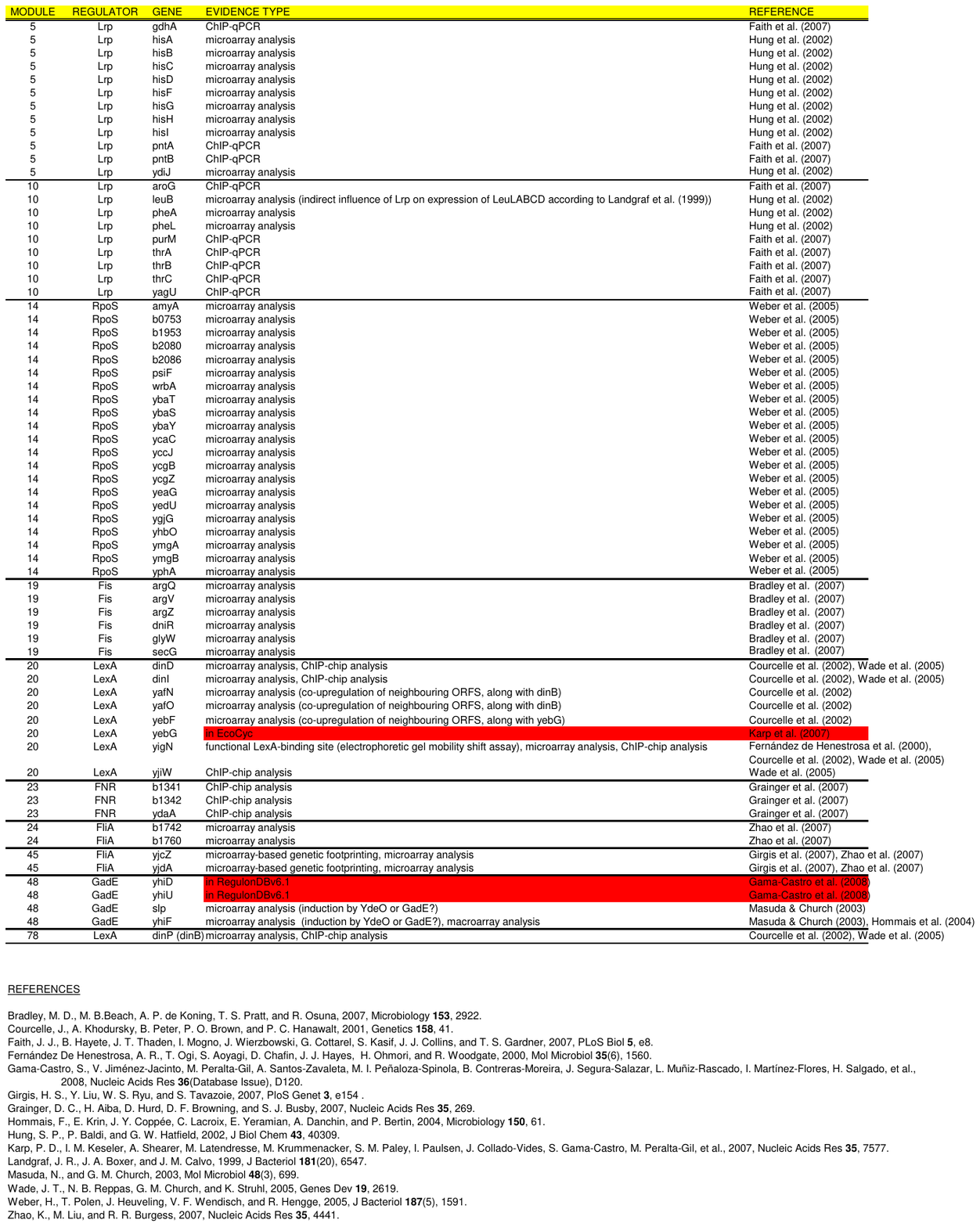}
  \label{tab:new_int}
  \caption{New interactions predicted in the 30\% precision LeMoNe network validated by 
    literature search.}
\end{table}

\newpage

\begin{figure}[ht!]
  \centering
  \includegraphics[width=\linewidth,viewport=0 250 595 780]{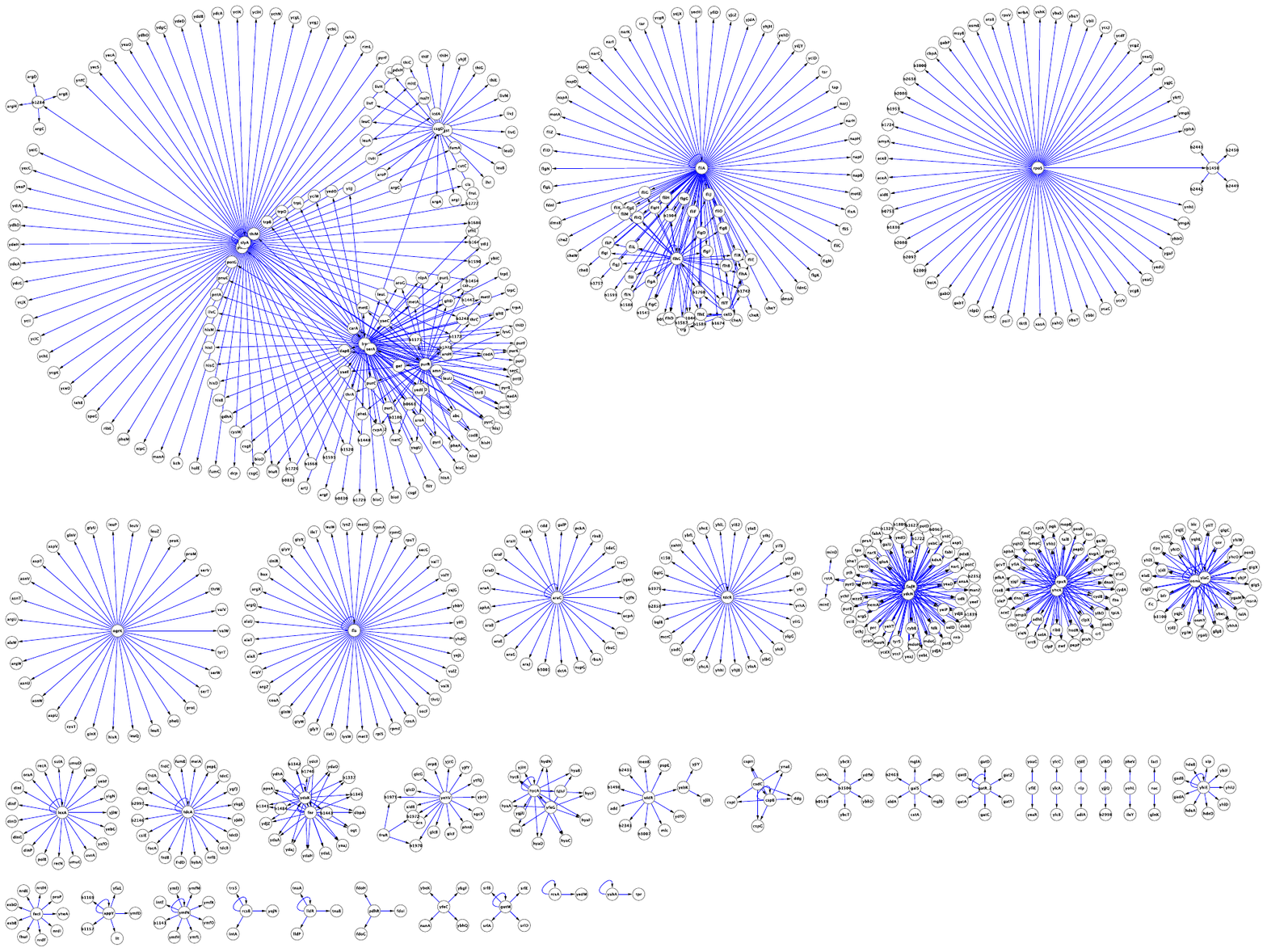}
  \caption{LeMoNe network for \textit{E. coli} at 30\% precision
    cutoff.}
  \label{fig:lemone-ecoli-network}
\end{figure}

\begin{figure}[ht!]
  \centering \includegraphics[width=\linewidth,viewport=0 200 595
  780]{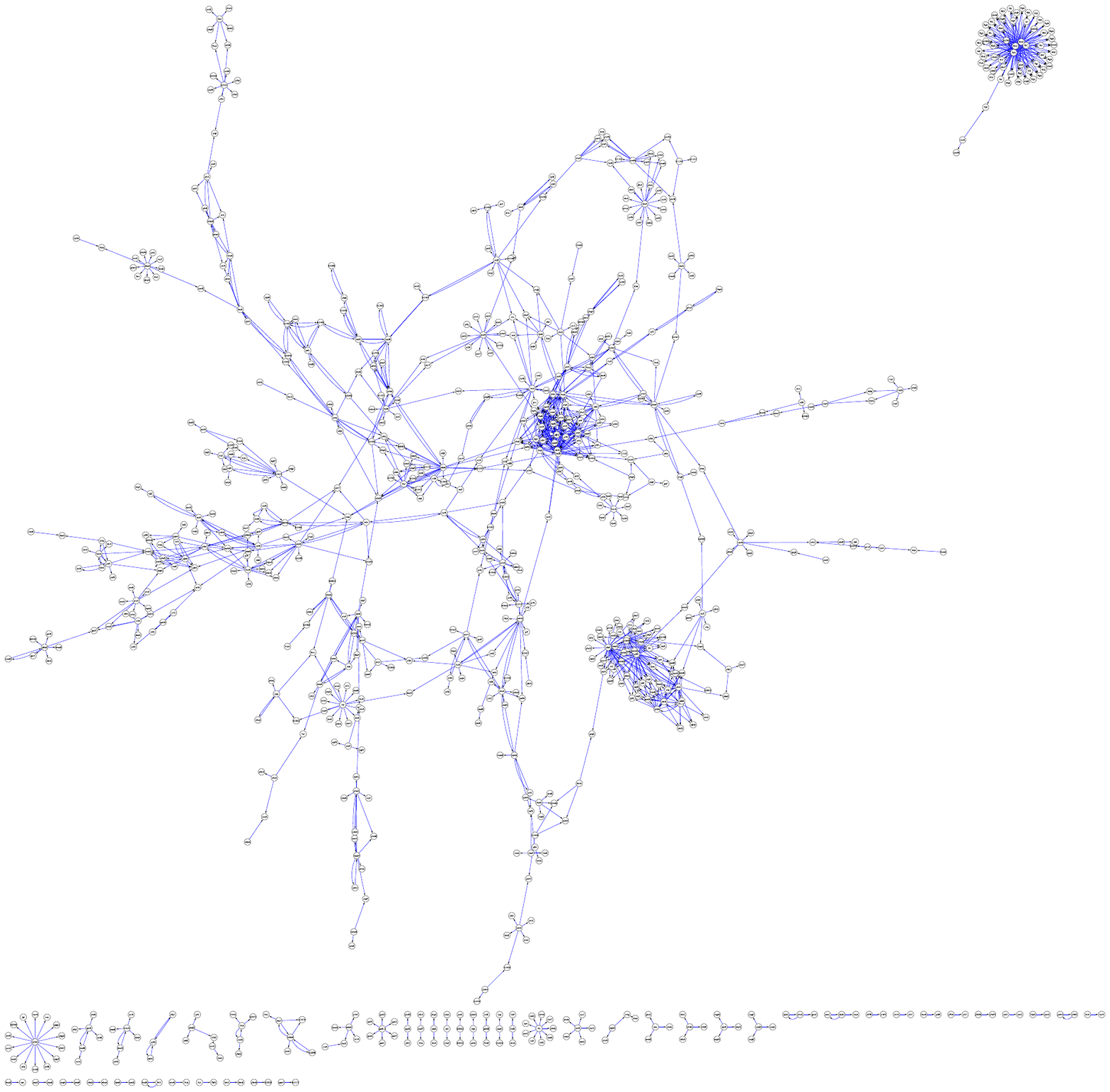}
  \caption{CLR network for \textit{E. coli} at 30\% precision
    cutoff.}
  \label{fig:clr-ecoli-network}
\end{figure}

\newpage

\begin{figure}[ht!]
  \centering
  \includegraphics[width=\linewidth,viewport=0 200 595 780]{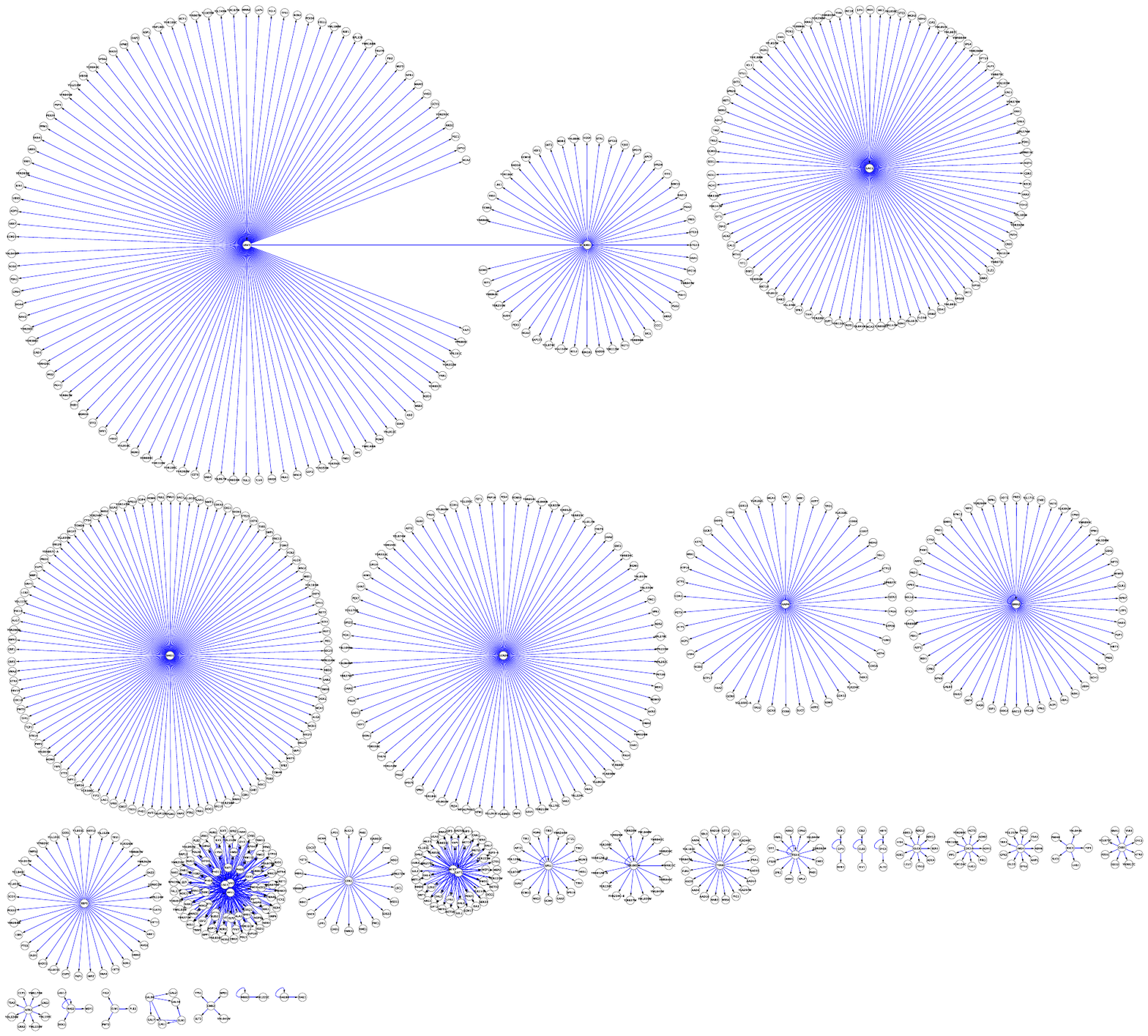}
  \caption{LeMoNe network for \textit{S. cerevisiae} at first 1070 predictions.}
  \label{fig:lemone-yeast-network}
\end{figure}

\newpage

\begin{figure}[ht!]
  \centering
  \includegraphics[width=\linewidth,viewport=0 200 595 780]{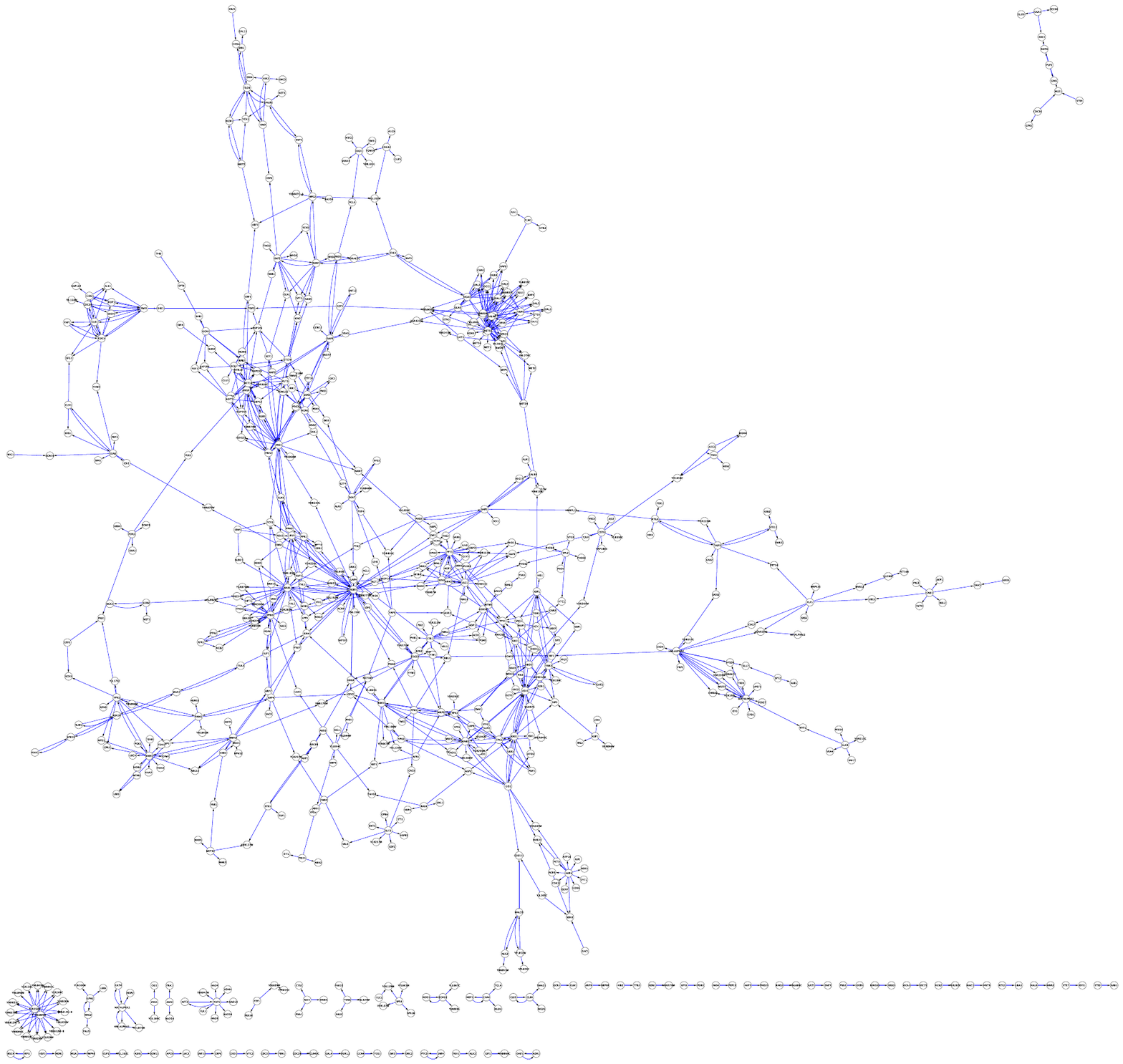}
  \caption{CLR network for \textit{S. cerevisiae} at first 1070 predictions.}
  \label{fig:clr-yeast-network}
\end{figure}

\newpage

\begin{figure}[ht!]
  \centering
  \includegraphics[width=0.4\textwidth]{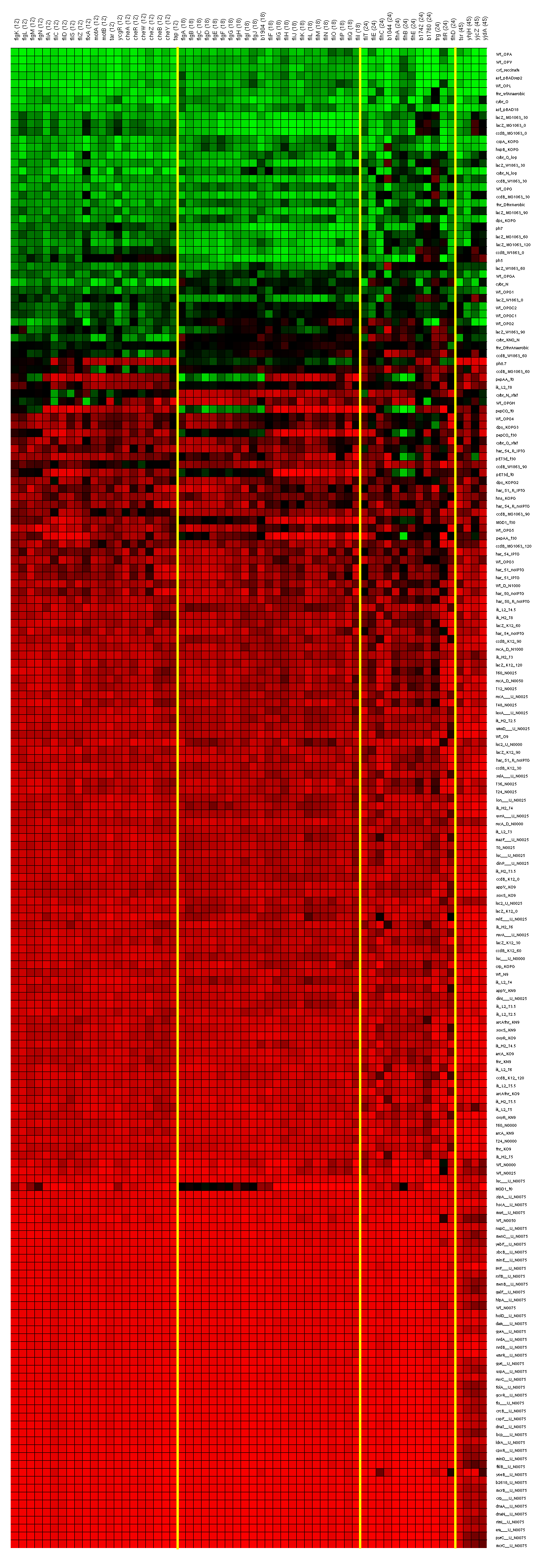}
  \caption{Expression levels of the genes in modules 12, 18, 24 and 45
    for \textit{E. coli}, with genes sorted in the same order as in
    Figure 5 (b). Conditions are sorted by the mean expression over
    all genes and yellow lines indicate module boundaries.}
  \label{fig:flagella_all}
\end{figure}

\end{bmcformat}
\end{document}